\documentclass[lettersize,journal]{IEEEtran}
\usepackage{amsmath,amsfonts}
\usepackage{algorithmic}
\usepackage{algorithm}
\usepackage{array}
\usepackage[caption=false,font=normalsize,labelfont=sf,textfont=sf]{subfig}
\usepackage{textcomp}
\usepackage{stfloats}
\usepackage{url}
\usepackage{verbatim}
\usepackage{graphicx}
\usepackage{cite}
\usepackage{xspace}
\usepackage{multirow}
\usepackage{tcolorbox}
\usepackage{xcolor}
\usepackage{pifont}
\usepackage{tikz}
\usepackage{booktabs}
\usepackage{makecell}
\usepackage[table]{xcolor}
\definecolor{lightgreen}{RGB}{210, 250, 210}
\definecolor{lightyellow}{RGB}{255, 255, 204}
\definecolor{lightred}{RGB}{255, 204, 204}
\usepackage{adjustbox}

\newcommand{\sol}{{\text{RECTor}}\xspace}
\newcommand{\Cor}{{\text{DeepCorr}}\xspace}
\newcommand{\Fea}{{\text{DeepCOFFEA}}\xspace}
\newcommand{\Tracker}{{\text{FlowTracker}}\xspace}

\newcommand{\CircledGOne}{
    \tikz[baseline=(char.base)]{
        \node[shape=circle, draw=black, fill=black, inner sep=1pt] (char){\textcolor{white}{\scriptsize{G$_1$}}};
    }
}

\newcommand{\CircledGTwo}{
    \tikz[baseline=(char.base)]{
        \node[shape=circle, draw=black, fill=black, inner sep=1pt](char){\textcolor{white}{\scriptsize{G$_2$}}};
    }
}

\begin{document}

\title{RECTor: Robust and Efficient Correlation Attack on Tor}

\author{Binghui Wu, Dinil Mon Divakaran, Levente Csikor, Mohan Gurusamy
        \thanks{This manuscript has been accepted for publication in IEEE Communications Magazine. The final version may differ from this preprint.}

}




\maketitle

\begin{abstract}
Tor is a widely used anonymity network that conceals user identities by routing traffic through encrypted relays, yet it remains vulnerable to traffic correlation attacks that deanonymize users by matching patterns in ingress and egress traffic. However, existing correlation methods suffer from two major limitations: limited robustness to noise and partial observations, and poor scalability due to computationally expensive pairwise matching. To address these challenges, we propose \sol, a machine learning-based framework for traffic correlation under realistic conditions. \sol employs attention-based Multiple Instance Learning (MIL) and GRU-based temporal encoding to extract robust flow representations, even when traffic data is incomplete or obfuscated. These embeddings are mapped into a shared space via a Siamese network, and efficiently matched using approximate nearest neighbor (aNN) search. {Empirical evaluations show that \sol outperforms state-of-the-art baselines such as \Cor, \Fea, and \Tracker---achieving up to 60\% higher true positive rates under high-noise conditions, and reducing training and inference time by over 50\%. Moreover, \sol demonstrates strong scalability: inference cost grows near-linearly as the number of flows increases.}
These findings reveal critical vulnerabilities in Tor’s anonymity model and highlight the need for advanced model-aware defenses.

\end{abstract}

\begin{IEEEkeywords}
Tor, correlation attack, privacy, machine learning, multiple instance learning, traffic analysis.
\end{IEEEkeywords}

\section{Introduction}
\label{sec:introduction}

With the increasingly surveilled and expanding landscape, anonymous communication systems like Tor~\cite{Tor-network-2004} have become essential for preserving user confidentiality and circumventing censorship. 
Tor provides client-side anonymity by enabling users to access the Internet without directly disclosing their IP addresses to destination servers. This is achieved by routing traffic through multiple encrypted relays, ensuring that no single relay simultaneously observes both the user's identity and destination. 
However, despite these robust safeguards, Tor remains vulnerable to \textit{traffic correlation attacks}, wherein adversaries correlate patterns between entry guard (having user's identity) and exit nodes (aware of the destination).
Such correlation can critically undermine user privacy, particularly when deployed at scale by state-level actors, compromised Internet providers, or malicious relay operators.
The deanonymization of Tor through traffic correlation analysis has received significant research attention. Such efforts extend to other forms of traffic analysis, including deep learning-based Keyword Fingerprinting (KF) attacks that infer specific user search queries from Tor traffic metadata~\cite{AsiaCCS2025}.
However, existing correlation {methods~\cite{Deepcorr-2018-CCS,Deepcoffea-2022-sp,Computer_sec_2023flowtracker,SubsetSum-2024-NDSS,Mixmatch-2024-PETS,AsiaCCS2025}} exhibit two primary limitations. 

(i)~\textbf{Robustness against noisy and partial observations: \CircledGOne} Many existing approaches rely on assumptions that often fail to hold in real-world deployment scenarios. Specifically, prior methods typically assume complete capture of all ingress (entry) and egress (exit) packets, from handshake initiation to session termination, and that each ingress flow has a perfectly corresponding egress flow~\cite{Deepcorr-2018-CCS,Deepcoffea-2022-sp,Computer_sec_2023flowtracker}. {FlowTracker~\cite{Computer_sec_2023flowtracker} introduces a relaxed setting by allowing one-to-many matching (i.e., an egress flow matched to a set of ingress candidates), and yet, it assumes that a correct match always exists.}

However, in realistic network deployments, adversaries (e.g., ISPs or malicious relays) often have only partial visibility into Tor traffic, {with access to only a subset of entry guards or exit nodes. This leads to incomplete or noisy observations, which means that in many cases, a matching ingress or egress flow simply does not exist. Moreover, adversaries may not capture an entire flow, missing early or late packets, which undermines early-segment correlation approaches~\cite{Deepcorr-2018-CCS,Deepcoffea-2022-sp}.} Real-world network conditions, such as packet delay, path congestion, user behavior diversity, and overlapping circuits, introduce significant noise and variability, which further degrade the reliability of traditional correlation strategies. {This necessitates robust representation learning techniques that can operate effectively under partial and noisy observation.}

{While FlowTracker also considers unrelated flows during matching, it does not address scenarios involving partial observation (e.g., missing packets) or structurally similar Tor flows that lack a true matching.
In our setting, ``noise” refers to unrelated Tor traffic that is structurally similar to the target flow, for example, different users visiting the same website, or the same user visiting different websites using the same Tor circuit. These realistic but confounding cases are not explicitly handled by prior works~\cite{Deepcorr-2018-CCS,Deepcoffea-2022-sp,Computer_sec_2023flowtracker,SubsetSum-2024-NDSS,Mixmatch-2024-PETS}.}

{(ii)~\textbf{Inefficient techniques:}\CircledGTwo} Existing correlation methods are computationally inefficient. Considering that typical Tor circuits persist for around 600 seconds, {which may result in the transmission of thousands of packets within a single Tor connection}, analyzing all packets from each flow becomes both time-consuming and resource-intensive. Current deep learning (DL)-based solutions, such as DeepCorr~\cite{Deepcorr-2018-CCS} and DeepCOFFEA~\cite{Deepcoffea-2022-sp}, typically rely on analyzing the first 300 to 800 packets to extract meaningful representations. This introduces significant {computational overhead and inefficiency. Moreover, these models degrade performance in terms of correlation accuracy when early packets are missing.} Their architectures, primarily built on convolutional neural networks (CNNs), are also limited in their ability to capture long-range sequential dependencies in traffic.

{In terms of computational complexity, prior methods~\cite{Deepcorr-2018-CCS,Deepcoffea-2022-sp,Computer_sec_2023flowtracker,SubsetSum-2024-NDSS} rely on exhaustive pairwise comparisons between every ingress and egress flow. That is, given $N$ ingress and $N$ egress flows, they must compute $\mathcal{O}(N^2)$ similarity scores---a process we refer to as \textit{pairwise} matching.}
Additionally, such quadratic complexity becomes particularly problematic under \textit{noisy} conditions, when only a small fraction (e.g., 10\%) of observed flows actually correspond to true matches. Efficiently filtering out {unrelated flows} is therefore essential for improving scalability and inference time in large-scale correlation attacks.

To address these challenges, we introduce \textbf{\sol}, a novel DL-based framework explicitly tailored for robust and efficient Tor traffic correlation under noisy and partial-mapping scenarios. 
\ding{172}~Unlike existing approaches that assume complete ingress-egress mappings, \sol adopts a more realistic threat model that includes partial traffic observation and heavy background noise. 
\ding{173}~Rather than focusing only on the first $n$ packets (as in previous works), \sol automatically learns the most relevant segments, thus representing traffic in a robust way even when (say, initial) packets are missing.
This is achieved by attention-based Multiple Instance Learning (MIL)~\cite{attention-MIL-2018} in combination with window-based flow segmentation.
\ding{174}~To improve efficiency, \sol learns nuanced temporal and statistical traffic patterns throughout the entire flow duration by integrating lightweight neural architectures, including Gated Recurrent Unit (GRU)~\cite{GRU-paper} encoders and attention mechanisms that select the most relevant parts.
Additionally, \sol employs a Siamese network~\cite{Siamese-2015} to embed traffic flows into a learned metric space. 
During training, the Siamese network brings correlated pairs closer together in the embedding space while pushing unrelated pairs apart. 
\ding{175}~Building on these learned embeddings, \sol applies approximate nearest neighbor (aNN) clustering to efficiently identify likely candidate matches, replacing the naive pairwise approach used in previous works.
Consequently, the computational complexity is reduced from quadratic to near-linear~(Section~\ref{section:clustering}), supporting large-scale correlation under realistic conditions. 
Table~\ref{tab:comparison} highlights the improvements of \sol in both robustness and scalability. 

We make the source code implementations of the models openly available to facilitate research\footnote{Code is available at: \texttt{\url{https://github.com/Binghui99/RECTor}}.}. This includes the implementation of the \sol framework and the data processing methods.

\begin{table*}[htbp]
\centering
\caption{{Comparison between \sol and prior correlation methods.}}
\label{tab:comparison}
\resizebox{1.6\columnwidth}{!}{
\begin{tabular}{lcccc}
\toprule
\textbf{Feature} & \textbf{DeepCorr}~\cite{Deepcorr-2018-CCS} & \textbf{DeepCOFFEA}~\cite{Deepcoffea-2022-sp} & \textbf{FlowTracker}~\cite{Computer_sec_2023flowtracker} & \textbf{\sol (ours)} \\
\midrule

\textbf{\makecell[l]{Partial Mapping Support}} &
\cellcolor{lightred}No &
\cellcolor{lightred}No &
\cellcolor{lightred}No &
\cellcolor{lightgreen}Yes \\

\textbf{\makecell[l]{Robustness to \\ Missing Packets}} &
\cellcolor{lightred}No &
\cellcolor{lightred}No &
\cellcolor{lightgreen}Yes  &
\cellcolor{lightgreen}Yes  \\

\textbf{Background Noise} &
\cellcolor{lightred}Fully mapped  &
\cellcolor{lightred}Fully mapped &
\cellcolor{lightyellow}Uncorrelated &
\cellcolor{lightgreen}Same website/circuit \\

\textbf{Model Architecture} &
\cellcolor{lightred}CNN+FC layers &
\cellcolor{lightyellow}CNN+GRU &
\cellcolor{lightyellow}CNN+stats &
\cellcolor{lightgreen}GRU+MIL+attention \\

\textbf{Similarity Measure} &
\cellcolor{lightred}Raw feature &
\cellcolor{lightyellow}Embedding &
\cellcolor{lightyellow}Feature distance &
\cellcolor{lightgreen}Embedding\&cluster \\

\textbf{Search Complexity} &
\cellcolor{lightyellow}Pairwise  $\mathcal{O}(N^2)$ &
\cellcolor{lightyellow}Pairwise  $\mathcal{O}(N^2)$ &
\cellcolor{lightyellow}Pairwise  $\mathcal{O}(N^2)$ &
\cellcolor{lightgreen}$\mathcal{O}(N \log N)$ \\

\bottomrule
\end{tabular}
}
\end{table*}

\section{Preliminaries, Challenges and Threat model}

\subsection{Anatomy of the Tor Network}
\begin{figure*}[!htbp]
    \centering
    \includegraphics[width=0.85\linewidth]{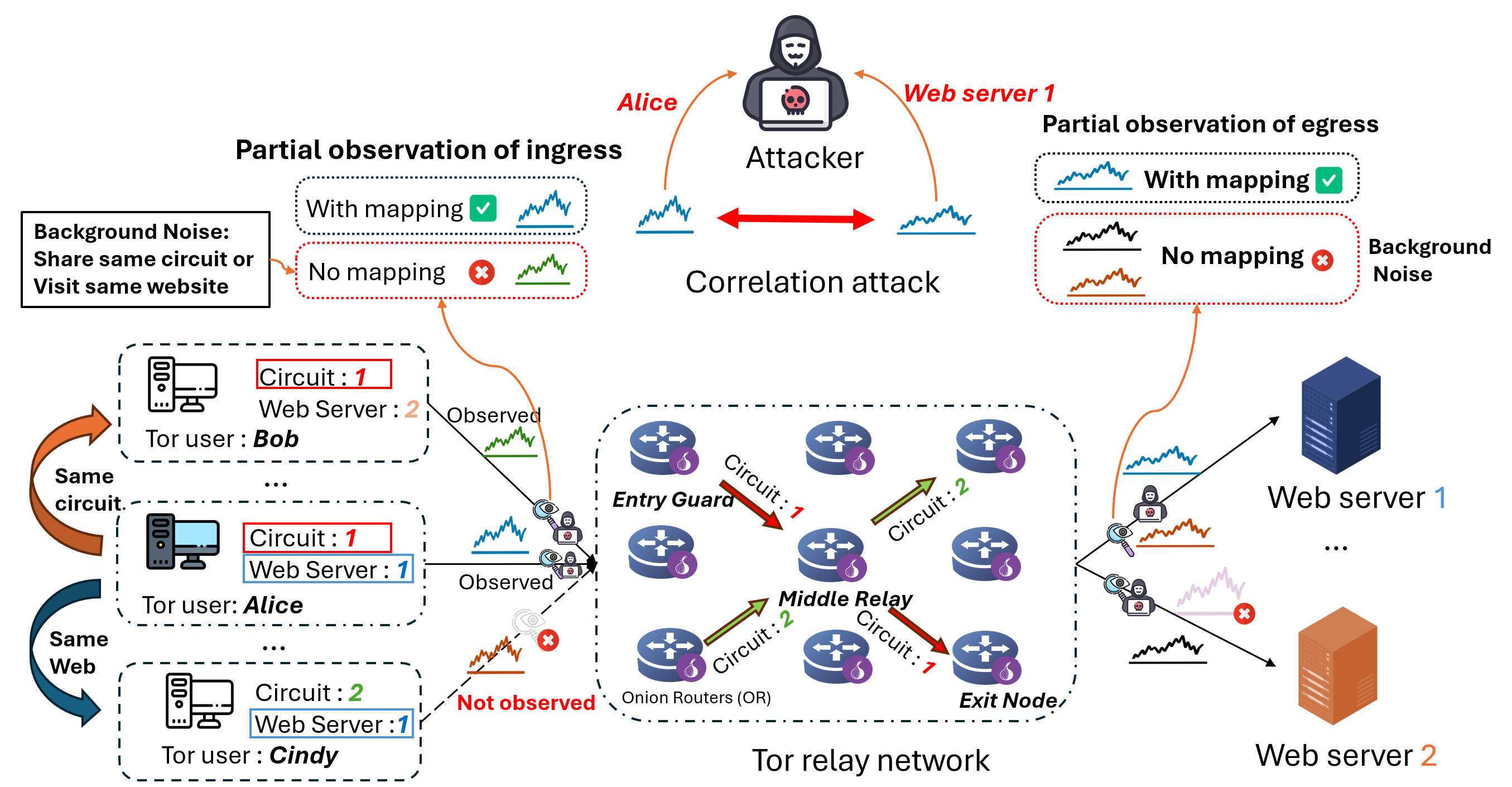} 
    \caption{{Tor network architecture and illustration of the threat model for traffic correlation attacks}}
    \label{fig:Threat-model}
\end{figure*}

The Tor network is composed of volunteer-operated relays known as \emph{onion routers} (ORs). 
Each OR publishes a \emph{descriptor} containing its public keys, network addresses, and other metadata, which are collectively stored in a global \emph{directory}. 
When a client wishes to connect anonymously to a destination server through Tor, it employs an \emph{Onion Proxy} (OP) that obtains relay information from this directory and constructs a multi-hop \emph{circuit}. 
By default, a circuit consists of three relays: an \emph{entry guard}, a \emph{middle relay}, and an \emph{exit node}~\cite{traffic-analysis-on-Tor}. {Figure~\ref{fig:Threat-model} illustrates the layered encryption of Tor and how vantage points can observe traffic at entry and exit nodes.}
{Throughout this article, we define a flow as a unidirectional or bidirectional sequence of packets identified by a 5-tuple: (source IP, destination IP, source port, destination port, protocol). Each flow corresponds to a user session (e.g., a website visit) over the Tor network. The term traffic is used to refer to aggregate packet-level activity, including all flows across users, circuits, or vantage points.}

\paragraph{Entry, Middle, and Exit} The \emph{entry guard} is the first hop in the Tor circuit that the client connects to directly.
It forwards traffic to the \emph{middle relay}, which subsequently passes it on to the \emph{exit relay}. 
Finally, the exit relay connects to the destination server on behalf of the client. 
Each pair of relays exchanges data through a dedicated TLS connection, referred to as a \emph{channel}, while the overall end-to-end flow of data from client to server is called a \emph{stream}.

\paragraph{Layered Encryption} Tor’s layered encryption model, often compared to the layers of an onion, ensures that each relay decrypts only one layer of ciphertext using its own key. 
Consequently, each relay knows only the previous and next hops in the circuit without learning the actual message content or the true identities of either end. 
This design prevents a single relay from linking the client to the destination, significantly enhancing user anonymity.

\paragraph{Circuit Rotation and Performance} To further reduce the risk of traffic analysis, Tor periodically (usually every ten minutes) discards old circuits and builds new ones. 
Although routing traffic through multiple volunteer relays worldwide may introduce additional latency, this is generally considered an acceptable trade-off for improved anonymity. 
Users should still exercise caution when entering personal credentials on external sites since once data leaves the Tor network at the exit relay, the final destination may observe or record it.

\subsection{Flow Correlation Attacks}

{\bf Flow correlation attacks} target to de-anonymize Tor (and other anonymity networks). 
By correlating an incoming Tor flow at the entry guard with an outgoing flow at the exit node, adversaries can reveal the true IP addresses of communication endpoints or, inversely, link a source to its destination. 
It is concerning for specific individuals, e.g., activists/journalists operating under authoritative regimes, who are very sensitive to their Internet activities being leaked to adversaries.
To launch such an attack, the adversary needs to observe both such flows. State-of-the-art network traffic analysis increasingly represents flows through learned embeddings and determines similarity via distance-based metrics in the embedding space~\cite{ZEST-2024,Amoeba-2023-CoNext,PANTS-2025-Usenix,2025-fastflow,UniNet-2025-TCCN}.
Analogously, traffic correlation methods typically operate by aligning low-level flow features, including inter-packet timing and throughput patterns.

There are two main attack strategies to conduct flow correlation attacks: i)~\emph{Malicious relays:} An adversary can operate a substantial number of malicious Tor relays~\cite{tan2022anonymity-TON}. 
By monitoring and recording traffic \emph{inside} the Tor network, they gather high-resolution data about packet dynamics, potentially enabling accurate flow correlation. 
ii)~\emph{Compromised vantage points:} Alternatively, an adversary can operate infrastructure-level entities such as Autonomous Systems (ASes) or Internet Exchange Points (IXPs)~\cite{2024-Survey-Tor} to intercept large portions of Tor traffic in transit, facilitating end-to-end correlation.

\subsection{Threat Model}\label{sec:threat_model}

We aim to evaluate flow correlation attacks under {practical and adversarially challenging} conditions, focusing on both effectiveness and scalability. {Our threat model approximates} a global passive adversary with vantage points at entry guards and exit nodes, {similar to prior work but with relaxed assumptions on visibility.} These vantage points may be operated by hostile relays, compromised ISPs, or surveillance entities controlling key network junctures. 
While the term “global” typically suggests omnipresent observation capabilities, in practice, such coverage is rarely achieved. Traffic collection is often incomplete due to infrastructure limitations, jurisdictional boundaries, or selective deployment of surveillance infrastructure. 
As a result, adversaries may only monitor a subset of ingress and egress points, leading to partial observations. This imperfect visibility complicates the correlation task and better reflects real-world constraints compared to fully observable settings assumed in earlier work.
Furthermore, under such partial visibility, the adversary must contend with “noisy” background traffic, i.e., unrelated Tor flows generated by other users. This noise can significantly hinder correlation accuracy, especially when the interfering traffic originates from the same website or uses the same Tor circuit structure, thereby exhibiting highly similar traffic patterns to the true matches.

The adversary observes a collection of $N$ incoming and $M$ outgoing flows—traffic entering the Tor network via entry guards and exiting through exit nodes. 
Unlike prior works~\cite{Deepcorr-2018-CCS,Deepcoffea-2022-sp,Computer_sec_2023flowtracker,SubsetSum-2024-NDSS,Mixmatch-2024-PETS}, which assume complete and clean ingress–egress mappings, we adopt a more realistic threat model that accounts for partial and noisy observations---where many flows have no true counterpart and background traffic may closely resemble valid matches due to shared websites or circuits.
Only a fraction of the observed flows, denoted by the \textit{mapping ratio} $\sigma \in [0, 1]$, represents valid ingress-egress pairs associated with the same user session. {The remaining flows, i.e., unmatched ingress or egress traffic, constitute background noise.}
{This noise is not merely random or uncorrelated. It often arises from users visiting the same website or using overlapping circuits, making the unrelated flows structurally similar to the true matches. Such fine-grained similarity has not been adequately handled by prior works~\cite{Deepcorr-2018-CCS,Deepcoffea-2022-sp,Computer_sec_2023flowtracker,SubsetSum-2024-NDSS}.}

\begin{figure*}[htbp]
    \centering
    \includegraphics[width=0.65\linewidth]{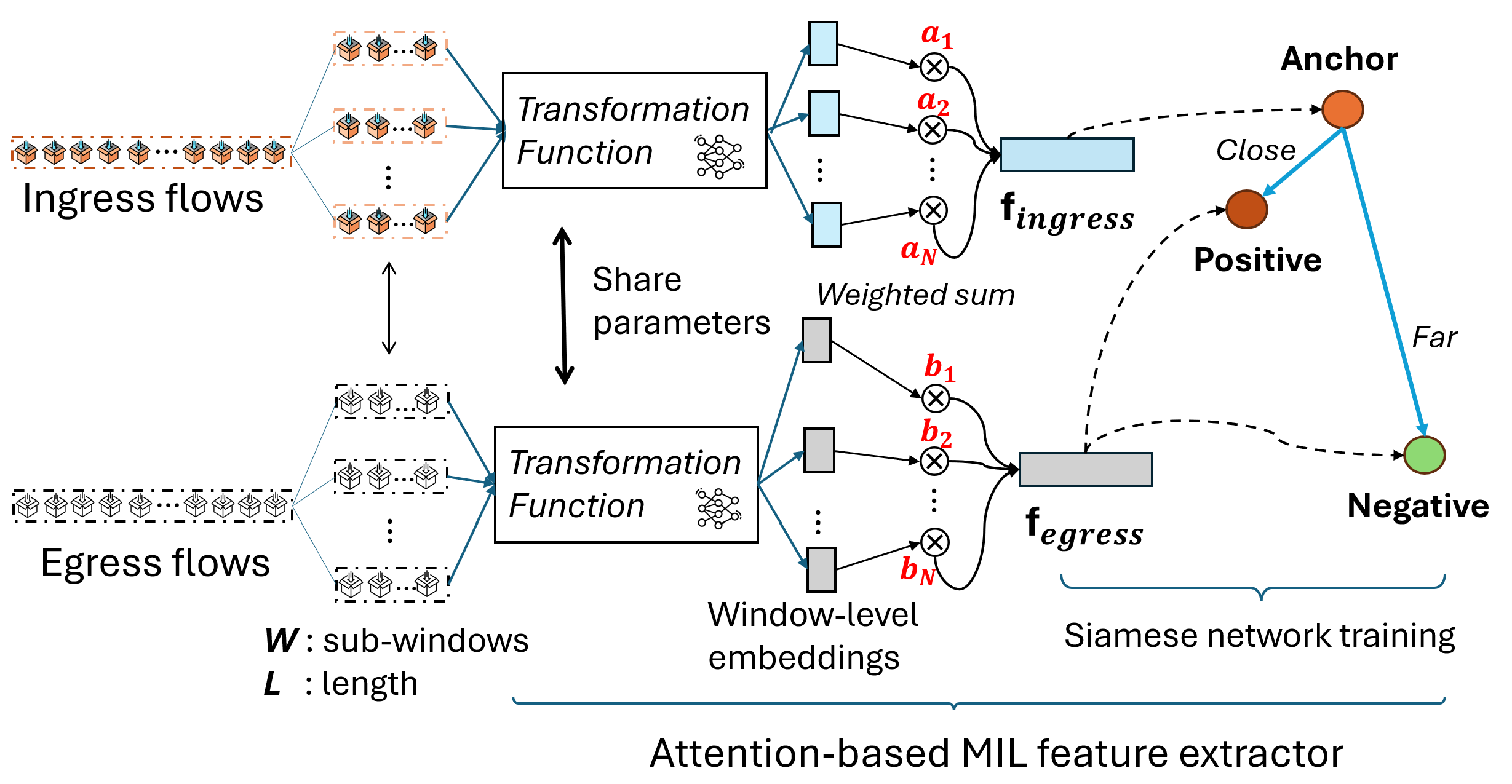} 
    \caption{Window partition and training of feature extractor in \sol}
    \label{fig:FNE}
    \vspace{-0.3cm}
\end{figure*}

As a result, identifying which ingress and egress flows truly correspond becomes a significantly more challenging task than traditional one-to-one matching. The adversary’s goal is to correlate flows accurately while filtering out noise and minimizing false matches. Since Tor encrypts all traffic, payload inspection is infeasible. However, observable side-channel features---such as packet size sequences, inter-arrival timing, and flow durations---remain accessible even under traffic obfuscation. {These subtle features are further complicated by padding strategies or shared browsing patterns across users.}

{To reflect practical deployment constraints, we also assume that the adversary’s computational resources are limited. Exhaustive pairwise comparison between all $N \times M$ flow pairs is not scalable in large-scale settings.} Hence, any effective attack must balance accuracy with efficiency, a core motivation for our design (see Figure~\ref{fig:Threat-model}).
We consider two operational scenarios that differ by mapping ratio $\sigma$:

\paragraph{Fully Mapping Scenario}
In this idealized case, $\sigma = 1$, meaning every ingress flow has a corresponding egress flow. {This setting aligns with the assumptions in DeepCOFFEA~\cite{Deepcoffea-2022-sp} and FlowTracker~\cite{Computer_sec_2023flowtracker}, where each flow is known to have a valid match within the candidate set.} With no background noise, correlation becomes a simpler classification problem, albeit still requiring many comparisons.

\paragraph{Partially Mapping Scenario}
In more realistic environments, $\sigma < 1$, meaning many ingress or egress flows lack any matching counterpart. {This introduces high volumes of structurally similar but unmatchable traffic that act as confounding noise.} {These unmatched flows may originate from the same destination websites or even be routed over the same Tor circuits, further increasing false positive risks.} Efficient and robust correlation under this scenario is a central goal of our framework, requiring noise-tolerant representation learning and scalable matching techniques.

\section{\sol Architecture}\label{sec:framework}

In this section, we present the architecture of \sol, {a scalable and noise-resilient framework for Tor traffic correlation under partial observations and challenging adversarial conditions.} 
{Our design directly addresses the two main challenges identified in Section~\ref{sec:introduction} : i)~noisy and incomplete traffic mappings, and ii)~the computational inefficiency of existing correlation methods.}
\sol comprises three key components: 
i)~{flow preprocessing and window segmentation for handling incomplete flows}; 
ii)~a robust feature extractor based on attention-based MIL; and 
iii)~an efficient candidate search module using aNN matching.
{Components (i) and (ii) work together to address}\CircledGOne (robust correlation under partial and noisy observation), as illustrated in Figure~\ref{fig:FNE}; 
{while component (iii) reduces runtime cost and scales the attack to large datasets}, addressing\CircledGTwo (computational efficiency), as shown in Figure~\ref{fig:deplo}. 
Below, we describe each module in detail, highlighting how they contribute to \sol’s performance.

\subsection{Data Processing and Window Partition}
\label{sec:preprocessing}

To handle noisy and partial observations, each Tor flow is divided into $W$ non-overlapping windows of fixed time duration (Figure~\ref{fig:FNE}). 
{This segmentation exposes diverse patterns from the start, middle, and end of each flow, allowing attention to prioritize informative segments while downweighting noisy or incomplete parts.}
Within each window, we extract three feature types: 
i) packet size statistics (e.g., average, variance); 
ii) inter-arrival times (IATs) {reflecting timing behavior}; 
and iii) flow direction indicators distinguishing upstream vs. downstream. 
Missing or incomplete windows are zero-padded for consistency, producing a uniform input shape of $(W \times L \times 2)$, where $L$ is the max number of packets and $2$ represents bidirectional features.
{This structure ensures robustness, as errors or missing packets in one window minimally affect the overall representation.}

\subsection{Attention-based MIL for Flow Representation}\label{sec:embedding}

{Rather than processing each Tor flow as a long, monolithic sequence, we model it as a collection (“bag”) of smaller, fixed-size window segments. 
This enables the model to focus selectively on the most informative segments.} 
This design improves robustness in real-world conditions, where flows are often incomplete, truncated, or mixed with unrelated Tor traffic.
To achieve this, \sol adopts an attention-based MIL framework. Each window is first encoded using a 2-layer GRU, capturing local temporal dynamics. 
The resulting embeddings are passed to an attention layer, which assigns a learnable importance weight to each.
{These attention weights let the model emphasize windows that reveal traffic patterns—such as bursts, delays, or directional shifts, while reducing weight of noisy or irrelevant segments.} 
The final flow-level embedding is computed as a weighted sum over all window embeddings.
The end-to-end embedding pipeline is illustrated in Figure~\ref{fig:FNE}.
To match related flows, we use a Siamese neural network~\cite{Siamese-2015}. 
{The model is trained with triplets---each consisting of an anchor (e.g., an ingress flow), a positive match (the corresponding egress), and a negative example (an unrelated egress).} 
{It learns to minimize the distance between anchor-positive pairs and maximize it for anchor-negative pairs, enforcing a margin in the learned embedding space.} 
This training strategy {promotes tight clustering of correlated flows and separation from distractors}, even under noisy or partial observations.
Together, these components improve correlation accuracy under adversarial conditions~\CircledGOne.

\subsection{Efficient Candidate Selection}\label{section:clustering}

\begin{figure*}[!htbp]
    \centering
    \includegraphics[width=0.65\linewidth]{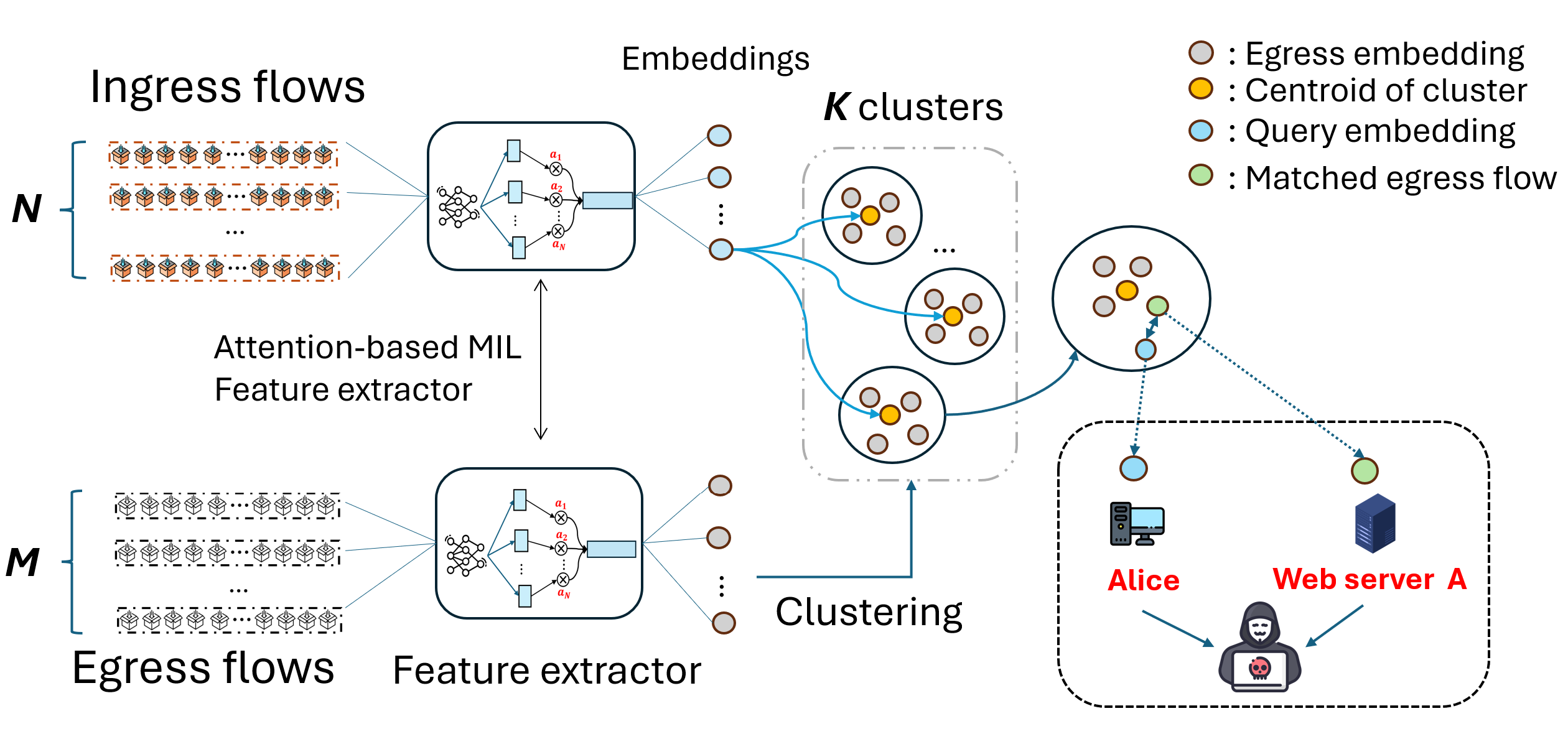} 
    \caption{Overview of efficient approximate search and matching in \sol}
    \label{fig:deplo}
\end{figure*}

Once robust embeddings are obtained, the next challenge is matching them efficiently at scale. 
For each candidate ingress–egress pair, we compute a matching score---the cosine similarity between their embeddings, and apply a decision threshold to determine whether they form a match.
{Exhaustively comparing each ingress flow against all $M$ egress flows incurs $\mathcal{O}(N \cdot M)$ complexity, which becomes prohibitive as traffic volume grows.} 
To address this, \sol adopts an aNN strategy~\cite{IVF-2010}. 
As illustrated in Figure~\ref{fig:deplo}, we apply $k$-means clustering to partition the egress embeddings into $K$ coarse clusters, each represented by a centroid and an inverted index of assigned flows~\CircledGTwo.

{To support scalable matching, all $M$ egress flows are grouped into a small number of clusters using $k$-means in the learned embedding space.}
Each flow is assigned to its nearest cluster centroid, which significantly narrows the search space for subsequent comparisons.
{For each incoming (ingress) flow, \sol first identifies a few clusters that are closest to it, typically 5 to 10, and only compares against the egress flows within those clusters.}
{This design cuts complexity from quadratic to near-linear. For each ingress flow, locating the nearest clusters requires logarithmic time, and only a small subset of flows within those clusters is compared. With an appropriate choice of cluster count, the overall cost scales as $\mathcal{O}(N \log N)$.}
This approach is particularly beneficial under partial mappings (i.e., low $\sigma$), where the majority of flows are unrelated background traffic. Instead of comparing every ingress-egress pair, \sol focuses computation on candidates within the nearest clusters in embedding space, which are most likely to contain the true matches. 
In practice, even modest clustering (e.g., 100 groups) enables near real-time inference while maintaining high recall and precision.

\section{Experimental Setup}\label{sec:experiments}

This section presents a carefully designed evaluation to demonstrate the performance and practicality of \sol. 
We consider two scenarios: fully mapped and partially mapped correlation tasks (see Section~\ref{sec:threat_model}). 
Our evaluation focuses on correlation accuracy, robustness to noise, and computational efficiency---measured via True Positive Rate (TPR), False Positive Rate (FPR), and inference time.

\subsection{Dataset and Preprocessing}
\label{sec:exp-setup}

We utilize the publicly available DeepCOFFEA dataset~\cite{Deepcoffea-2022-sp}. While the traffic in this dataset was collected in a controlled environment, it is emulated rather than simulated. 
Specifically, real applications (i.e., Tor browsers) are programmatically instructed (through automation) to visit certain websites (the top 100 as per Alexa's ranking) via the live Tor network using virtual machines. 
This focus on popular websites follows common practice in prior work~\cite{Deepcorr-2018-CCS,Deepcoffea-2022-sp,Computer_sec_2023flowtracker} and provides a realistic, widely adopted benchmark for Tor correlation evaluation.
We note that performing correlation attacks based entirely on unseen websites remains an important open problem for future research.
This approach captures realistic Tor behavior, including circuit construction, encryption patterns, and network jitter. 
Although collected in a controlled environment, the use of real-world Tor browsers ensures significant geographic and temporal diversity, since onion routers and exit nodes are globally distributed, and their actual locations can change every time a new Tor circuit is established (which happens frequently).
{A 60-second limit is set for page loading within this environment, functioning as a timeout to allow browsers sufficient time to fully load websites.}

Traffic is generated using 341 distinct Tor circuits, with each circuit visiting all 100 websites sequentially, yielding 34,100 website sessions---i.e., 341 circuits × 100 sessions. 
Each session generates a unique ingress–egress flow pair. We use 300 circuits (30,000 visits) for training, and hold out 41 circuits (4,100 visits) for testing. 
During testing, noise traffic is taken only from the 41 held-out circuits, with a fraction $\sigma$ of flows used as matched pairs and the remaining $(1-\sigma)$ as unmatched background flows.
This split follows a conventional 9:1 ratio commonly used in machine learning, ensuring sufficient training data while enabling evaluation on unseen circuits and browsing behaviors.
Each flow is divided into 10 non-overlapping 5-second windows. 
Within each window, we extract: 
i)~\textit{packet size} and \textit{inter-arrival time (IAT)} statistics, and 
ii)~\textit{flow direction}.
All windows are zero-padded to 100 packets for consistency.
{This window size balances detail and efficiency; larger values yielded diminishing returns.}


To emulate realistic noise, we introduce partial mapping conditions during testing with $N=M=500$ flows. 
Only a fraction $\sigma \in \{0.1, 0.3, 0.5, 0.8, 1.0\}$ are true ingress–egress matches; the rest act as {background unmatched flows (noise)}. 
Importantly, the unmatched flows are sampled from the same set of circuits and websites, making them structurally similar to genuine matches.
This setup simulates real-world Tor conditions where most observed flows lack counterparts.

\subsection{Baselines and Metrics}
\label{sec:baselines-metrics}
{Traditional machine learning models have shown limited effectiveness under realistic noise and partial traffic due to reliance on handcrafted features~\cite{Deepcorr-2018-CCS,Deepcoffea-2022-sp}. We therefore focus on recent DL-based methods.}
To contextualize our results, we compare \sol against three representative baselines:

\begin{itemize}
    \item \textbf{\Cor~\cite{Deepcorr-2018-CCS}}:
    One of the first CNN-based correlation approaches, \Cor processes the first $n$ packets of each flow to extract timing and size patterns via 1D convolutions. {It performs binary classification on flow pairs using a fully supervised, pairwise training setup.} While efficient, \Cor is vulnerable to missing-packet noise and does not scale well to partial observations.

    \item \textbf{\Fea~\cite{Deepcoffea-2022-sp}}:
    This method segments flows into fixed-size sub-windows, each independently matched using a voting scheme. {Though effective at reducing false positives, the design increases inference cost due to repeated comparisons.} \Fea represents a trade-off between improved correlation and higher runtime.

    \item \textbf{FlowTracker~\cite{Computer_sec_2023flowtracker}}:
    {A recent approach that aggregates flow-level features and uses an autoencoder to learn compact embeddings for matching. 
    It leverages fingerprint similarity and efficient ranking but operates under a closed-world assumption, i.e., it presumes that each egress flow has a true corresponding ingress candidate.}
\end{itemize}

We evaluate all methods on \emph{robustness}---measured via True Positive Rate (TPR) and False Positive Rate (FPR) under noisy and partially mapped conditions; and on {\textit{computational scalability}}, including training and inference performance. 
{All experiments were conducted on an Nvidia RTX 4090 GPU with 24GB memory to ensure consistent runtime comparison.}

\begin{table*}[htbp]
\caption{Efficiency and scalability comparison of traffic correlation methods.}
\label{tab:efficiency_combined}
\centering
\resizebox{1.76\columnwidth}{!}{%
\begin{tabular}{l|c|rrrrr||rrrrc}
\toprule
\multirow{2}{*}{\textbf{Method}} & 
\multirow{2}{*}{\textbf{Train (h)}} & 
\multicolumn{5}{c||}{\textbf{Inference Time (s) @ 500 Flows}} & 
\multicolumn{5}{c}{\textbf{Scalability: Inference Cost vs Flow Volume @ $\sigma{=}0.1$}} \\
& & $\sigma{=}0.1$ & 0.3 & 0.5 & 0.8 & 1.0 
  & 100 & 500 & 1000 & 2000 & Complexity (Big O notation) \\
\midrule
\textbf{DeepCorr}~\cite{Deepcorr-2018-CCS} & 60 
  & 26.82 & 26.83 & 27.34 & 27.71 & 28.09 
  & 1.0× & 5.0× & 10.1× & 20.4× & $\mathcal{O}(N^2)$ \\
\textbf{DeepCOFFEA}~\cite{Deepcoffea-2022-sp} & 72 
  & 0.2485 & 0.2495 & 0.2533 & 0.2564 & 0.2648 
  & 1.0× & 5.1× & 10.2× & 19.8× & $\mathcal{O}(N^2)$ \\
\textbf{FlowTracker}~\cite{Computer_sec_2023flowtracker} & 58 
  & 1.204 & 1.223 & 1.252 & 1.271 & 1.318 
  & 1.0× & 4.4× & 8.6× & 16.3× & \textbf{$\mathcal{O}(N^2)$} \\
\midrule
\textbf{\sol (ours)} & \textbf{23} 
  & \textbf{0.0060} & \textbf{0.0163} & \textbf{0.0344} & \textbf{0.0972} & \textbf{0.1191} 
  & \textbf{1.0×} & \textbf{1.4×} & \textbf{1.9×} & \textbf{2.7×} & \textbf{$\mathcal{O}(N \log N)$} \\
\bottomrule
\end{tabular}
}
\end{table*}

\begin{itemize} 

\item \textbf{TPR and FPR:} 
TPR reflects the ability to recover true matches, while FPR indicates how often unrelated flows are incorrectly linked.
In large-scale or noisy environments, a high FPR overwhelms the system with false correlations, leading to low utilization of such a solution (for the attacker). 
A high TPR is only meaningful when it can be achieved at a low FPR.
We vary the matching threshold across its full range, tracing out the different [TPR, FPR] trade-offs.

\item \textbf{{Scalability and Runtime:}} 
{Beyond raw inference time, we assess how each method scales under partial mappings and background noise.}
We measure inference time over 500 flows at different mapping ratios $\sigma$ to observe performance. 
{To evaluate scalability, we also track how computational complexity grows as the number of flow pairs increases from 100 to 2000.}

\end{itemize}

\section{Results and Discussion}\label{sec:results}

We evaluate the performance of \sol and baseline methods under varying noise levels, controlled via different mapping ratios ($\sigma$).
All results are averaged over five independent runs to ensure robustness and consistency.

\subsection{Robustness Analysis}\label{sec:robustness_analysis}

\begin{figure}[!htbp]
    \centering
    \includegraphics[width=0.95\linewidth]{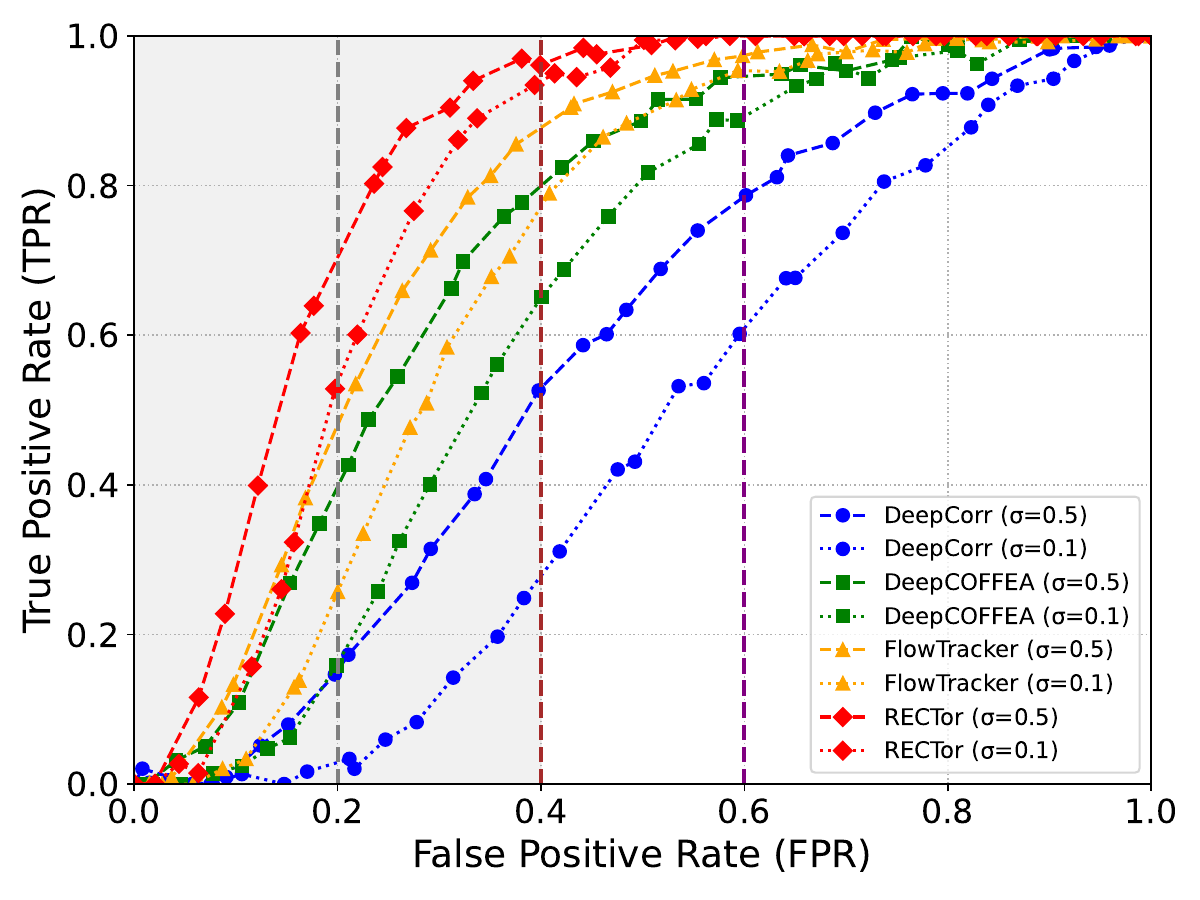}
    \caption{TPR vs. FPR comparison for RECTor and baselines under partial mapping ($\sigma=0.5$) and high noise ($\sigma=0.1$).}
    \label{fig:TPRVSFPR}
\end{figure}
In real-world Tor environments, adversaries typically operate under noisy and partially observable conditions. 
Robust traffic correlation methods must therefore remain effective even when only a subset of flows are relevant and the rest are misleading background traffic.
We evaluate \sol against state-of-the-art baselines, including \textbf{\Cor}, \textbf{\Fea}, and \textbf{\Tracker}, using TPR–FPR curves at two practical noise settings: {$\sigma = 0.5$ (moderate noise)} and $\sigma = 0.1$ (heavy background traffic). 
We omit the fully mapped $\sigma=1.0$ case to conserve space, as it is less representative of real deployment scenarios.

\paragraph*{Resilience to Background Noise}
Figure~\ref{fig:TPRVSFPR} TPR–FPR trade-off under moderate noise ($\sigma=0.5$) and high noise ($\sigma=0.1$).
\sol consistently achieves the best trade-off across conditions. At $\sigma = 0.5$, it reaches 85\% TPR at 20\% FPR, outperforming DeepCOFFEA and FlowTracker by large margins. Even under the challenging $\sigma = 0.1$ scenario, \sol maintains 70\% TPR at 20\% FPR, while baselines degrade significantly. {In such noisy regimes, even moderate gains significantly reduce computational burden by minimizing false leads while still identifying meaningful matches.}

{This performance gap caused by three architectural advantages. First, unlike DeepCorr and DeepCOFFEA, which statically process fixed initial flow segments, \sol leverages attention-based MIL to dynamically re-weight flow segments. This enables the model to ignore noisy or truncated windows, whether they occur at the beginning, middle, or end of a session.} 
{Second, FlowTracker assumes that every egress has a true matching ingress flow within a closed candidate set, making it prone to false positives when this assumption breaks down. \sol relaxes this constraint via a tunable mapping ratio~$\sigma$, enabling it to operate reliably in open-world or partially observable scenarios.}
{Third, \sol uses GRU-based encoders that preserve temporal ordering and long-range dependencies, resulting in flow embeddings that remain discriminative even when only partial traffic is observed.}

{The ability to preserve high TPR while suppressing FPR is essential in realistic settings, where false matches can mislead investigations or generate noise at scale.} These results confirm that \sol is not only more accurate than existing methods, but also significantly more resilient to real-world Tor noise and partial observability.

\subsection{Efficiency and Scalability Analysis}\label{sec:efficiency_analysis}
Efficiency plays a crucial role in making traffic correlation attacks feasible at scale. 
{Table~\ref{tab:efficiency_combined} provides a unified view of both per-instance runtime and scalability trends under varying mapping ratios ($\sigma$) and growing flow volumes.} 
\sol not only enhances accuracy but also significantly reduces training and inference time compared to baselines.

{On the left of the table, we compare training hours and inference time (in seconds) for each method on a batch of 500 flows. 
On the right, we show how inference time scales as the number of flows increases under high background noise ($\sigma = 0.1$), normalized to runtime at 100 flows. 
This upper-bound scenario reflects practical network conditions where true mappings are sparse, answering key reviewer concerns about performance under noise and partial observability.}

\subsubsection{Training Efficiency} 
Training time is measured until the loss reaches 0.004, following the settings in~\cite{Deepcoffea-2022-sp}. 
As shown in Table~\ref{tab:efficiency_combined}, \Cor takes 60 hours due to costly pairwise comparisons, while \Fea’s dual-model CNN-based feature extractor extends training to 72 hours. 
{FlowTracker improves on both, training in 58 hours by relaxing full pairwise search and using aggregated statistics.} 
In contrast, \sol completes training in just 23 hours, {the shortest among all methods}. It thanks to its lightweight GRU-based MIL architecture with attention aggregation.

\subsubsection{Inference Efficiency and Scalability} 
We evaluate inference time on 500 flow pairs across multiple mapping ratios ($\sigma \in \{0.1, 0.3, 0.5, 0.8, 1.0\}$). 
\Cor incurs the highest computational overhead, requiring 28.09s at $\sigma=1.0$ and 26.82s at $\sigma=0.1$, reflecting its $N^2$ pairwise structure. 
\Fea achieves better efficiency through voting and CNNs, dropping inference time to under 0.27s. 
{FlowTracker offers a middle ground at $\sim$1.2s per batch, but still scales quadratically with traffic volume.}

In contrast, \sol leverages GRU encoding, attention-based MIL, and aNN search to avoid exhaustive matching. 
This results in dramatically lower inference times---just 0.0060s at $\sigma=0.1$, a 99.8\% reduction from \Cor and 96\% from \Fea. 
{More importantly, as shown in the right-hand panel of Table~\ref{tab:efficiency_combined}, \sol maintains near-linear scalability. 
When the number of flows grows from 100 to 2000, \sol's inference cost rises only 2.7×, compared to 20× for \Cor. 
This confirms \sol's practical deployability in high-volume Tor networks with background noise.}

\subsection{Defending Strategy}

Countering modern correlation attacks like \sol demands rethinking legacy defenses, as existing techniques, though effective against earlier statistical methods, are insufficient against deep learning-based models that capture structural and temporal patterns. Tor's use of fixed-size cells (512 bytes) and minor timing obfuscation was not designed to counter neural embeddings trained over thousands of noisy sessions~\cite{Sp2023sok}.

{To defend against \sol, we need \textbf{learning-aware countermeasures} that disrupt correlation signals without severely degrading performance.
One direction is \textit{adversarial traffic morphing}~\cite{2009-traffic-morphing,BAP-2021-Usenix,Defences-2024-Infocom}, where packet timings, directions, or sizes are perturbed dynamically using lightweight padding or reshaping strategies that confuse deep models (i.e., GRU encoders and MIL in \sol).} Another promising line of defense is \textbf{adversarial example generation in network traffic}~\cite{Amoeba-2023-CoNext,PANTS-2025-Usenix} which selectively modifies flow characteristics to degrade model confidence or induce incorrect associations during aNN matching.
Recent work such as Amoeba [12] and MalCL~\cite{AAAI-2025} has shown that even small, targeted noise can prevent reliable correlation by neural networks.
More broadly, protocol-level defenses that mix or aggregate flows at entry guards or rendezvous points could suppress flow-level uniqueness and further impair embeddings.
These defenses require comprehensive study, which we see as an important future direction.

\section{Conclusion}

In this work, we presented \sol, a robust and scalable framework for traffic correlation attacks on Tor under realistic conditions, including partial observability and background noise. 
By integrating attention-based MIL, temporal-aware embeddings, and approximate nearest neighbor search, \sol achieves high TPR at a low FPR while maintaining near-linear inference complexity. 
Through extensive evaluation, we demonstrated that \sol consistently outperforms prior approaches such as DeepCorr, DeepCOFFEA, and {FlowTracker}, achieving up to 40\% higher TPR under noisy conditions and reducing inference time by up to 99\%. 
{Our scalability study further shows that \sol is practical for large-scale deployment, even with sparse signal scenarios.} These findings underscore the continued vulnerability of Tor to sophisticated correlation attacks.

\section*{Acknowledgment}

This research/project is supported by the National Research Foundation, Singapore, and the Cyber Security Agency of Singapore under the National Cybersecurity R\&D Programme and the CyberSG R\&D Programme Office (Award CRPO-GC2-ASTAR-001). 
Any opinions, findings, conclusions, or recommendations expressed in these materials are those of the author(s) and do not reflect the views of the National Research Foundation, Singapore, the Cyber Security Agency of Singapore, or the CyberSG R\&D Programme Office. This research/project is also supported by the National University of Singapore (NUS).

\bibliographystyle{IEEEtran}
\bibliography{ref}

\begin{IEEEbiographynophoto}{Binghui Wu}
(binghuiwu@u.nus.edu) is a Ph.D. candidate with the Department of Electrical and Computer Engineering, National University of Singapore.
\end{IEEEbiographynophoto}
\vspace{-1.5cm}

\begin{IEEEbiographynophoto}{Dinil Mon Divakaran}
(Senior Member, IEEE; dinil\_divakaran@a-star.edu.sg) is a Senior Principal Scientist at the A*STAR Institute for Infocomm Research (A*STAR I$^2$R), Singapore. He is also an Adjunct Assistant Professor at the School of Computing in the National University of Singapore. His research interests are network security, web security, and AI security.
\end{IEEEbiographynophoto}
\vspace{-1.5cm}
\begin{IEEEbiographynophoto}{Levente Csikor}
(cslev@cslev.vip) is a Senior Scientist at the A*STAR Institute for Infocomm Research (A*STAR I$^2$R), Singapore.
His research explores the intricate dimensions of security and privacy in next-generation networks.
\end{IEEEbiographynophoto} 
\vspace{-1.5cm}

\begin{IEEEbiographynophoto}{Mohan Gurusamy}
(Senior Member, IEEE; gmohan@nus.edu.sg) is an Associate Professor with the Department of Electrical and Computer Engineering, National University of Singapore.
\end{IEEEbiographynophoto}

\end{document}